\documentclass[aps,prl,twocolumn,groupedaddress]{revtex4} %submit to arXiv
\usepackage{graphicx}
\usepackage{bm}
\begin{document}

\title{Bose-Einstein Condensation of $^{88}$Sr Through Sympathetic Cooling with $^{87}$Sr}
\author{P. G. Mickelson, Y. N. Martinez de Escobar,  M. Yan, B. J. DeSalvo, and T. C. Killian}
%\date{}                                           % Activate to display a given date or no date
\affiliation{Rice University, Department of Physics and
Astronomy, Houston, Texas, 77251}

\date{\today}

\begin{abstract}

We report Bose-Einstein condensation of $^{88}$Sr, which has a small, negative $s$-wave scattering length ($a_{88}=-2$\,$a_0$). We overcome the poor evaporative cooling characteristics of this isotope by sympathetic cooling with $^{87}$Sr atoms. $^{87}$Sr is effective in this role in spite of the fact that it is a fermion because of the large ground state degeneracy arising from a nuclear spin of $I=9/2$, which reduces the impact of Pauli blocking of collisions. We observe a limited number of atoms in the condensate ($N_{max}\approx 10^4$) that  is consistent with the value of $a_{88}$ and the optical dipole trap parameters.
\end{abstract}

%\pacs{32.80.Pj}

\maketitle
Bose-Einstein condensation of $^{88}$Sr has been pursued for over a decade because of the promise of efficient laser-cooling  to high phase space density using the $(5s^2)^1S_0$-$(5s5p)^3P_1$ narrow intercombination line \cite{kii99} and loading of optical dipole traps  that operate at the magic wavelength for this transition \cite{iik00}. Recent interest in $^{88}$Sr has focused on long-coherence time interferometers \cite{fps06}, optical frequency standards \cite{lvm08,atk10}, and the existence of low-loss optical Feshbach resonances \cite{ctj05,ekk08}.
There has also been great interest generally in quantum degenerate gases of alkaline-earth metal atoms and atoms with similar electronic structure because of potential applications in quantum computing in optical lattices \cite{dby08,grd09, rjd09} and creation of novel quantum fluids \cite{hgr09}.
% recent spectacular advances in optical frequency standards
%\cite{ykk08}.

%88Sr is special because it is .most abundant Sr isotope and it is very Weakly %interacting is special...

Early attempts to evaporatively cool $^{88}$Sr to quantum degeneracy in an optical dipole trap \cite{iik00,fdp06} were not successful in spite of  initial phase space densities as high as $10^{-1}$, presumably because of a small elastic scattering cross section. This was confirmed by measurements of the scattering lengths of all strontium isotopes using photoassociative \cite{mms05,mmp08} and Fourier-transform \cite{skt10} spectroscopy of Sr$_2$ molecular potentials, which found that $a_{88}=-2$\,$a_0$, where $a_0=5.29 \times 10^{-11}$\,m is the Bohr radius. Here, we report  Bose-Einstein condensation (BEC) of $^{88}$Sr through sympathetic cooling with $^{87}$Sr.

Divalent atoms such as strontium and ytterbium \cite{fst09,fts07}
often possess a  large number of stable isotopes, which enables mass tuning
of the $s$-wave scattering length. For strontium, the stable isotopes and
abundances are $^{88}$Sr (82.6\%),$^{87}$Sr (7.0\%),$^{86}$Sr (9.9\%), and
$^{84}$Sr (0.6\%). In such systems, the likelihood of finding an isotope
with a scattering length that enables efficient evaporative cooling is very
high, as was recently demonstrated through the condensation of $^{84}$Sr
($a_{84}=123$\,$a_0$) \cite{sth09,mmy09}. For an isotope that has a poor
scattering length for evaporative cooling, there are also numerous
opportunities to find another isotope that can be used effectively
for sympathetic cooling. For $^{88}$Sr, the fermionic isotope $^{87}$Sr
is well-suited for this role. It has a large and positive $s$-wave
scattering length of $a_{87}=96$\,$a_0$ \cite{mmp08,skt10} which leads to
efficient thermalization and evaporation as long as the system is not highly
polarized. The inter-isotope scattering length is also reasonable,
$a_{88-87}=55$\,$a_0$ \cite{mmp08,skt10}, so that in a mixture, $^{88}$Sr
will be efficiently cooled (and evaporated) through collisions with $^{87}$Sr.

Bosons are normally used to sympathetically cool fermions \cite{tsm01}
rather than  the other way around, because identical fermions suffer from
collisional Pauli blocking, which reduces evaporation efficiency in the
quantum degenerate regime \cite{dji99}. We do not observe significant limitations due to Pauli blocking in the experiments reported here, and we suspect this is because $^{87}$Sr  has a
large nuclear spin ($I=9/2$) and ground state degeneracy. This suppresses
the Fermi temperature and allows $^{88}$Sr to be cooled to high phase space
density before Pauli blocking of $^{87}$Sr collisions becomes important.

\begin{figure}
\includegraphics[clip=true,keepaspectratio=true,width=3in]{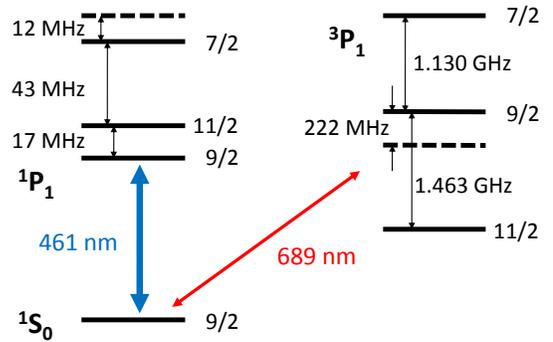}\\
\caption{(color online) Partial level diagram for $^{88}$Sr (- -) and $^{87}$Sr (---) showing hyperfine structure and isotope shifts \protect\cite{xld03,klj06,cqb05}. Total quantum number $F$ is indicated for $^{87}$Sr levels.}
 \label{Level Diagram}
\end{figure}

Details about our apparatus can be found in refs.\ \cite{mmp08,mma09,mmy09}.
Formation of ultracold mixtures of strontium isotopes benefits from the ability
to magnetically trap atoms in the metastable $(5s5p)^3P_2$ state
\cite{nsl03,xlh03,pdf05,fdp06}, which has a 10 min lifetime \cite{yka04}.
One isotope is trapped from a Zeeman slowed beam in a magneto-optical trap
(MOT) operating on the $(5s^2)^1S_0$-$(5s5p)^1P_1$ transition at 461 nm.
This transition is not closed and approximately 1 in $10^5$ excitations
results in an atom decaying through the $(5s4d)^1D_2$ state to the
$(5s5p)^3P_2$ state, where it can be trapped in the quadrupole magnetic
field of the MOT. After accumulating a desired number of atoms, limited by
the loading rate and observed $^3P_2$ lifetime of about 25\,s, the cooling laser frequency is then switched to cool and
accumulate another isotope. In our experiment,
we load $^{88}$Sr for 3 s and then $^{87}$Sr \cite{xld03} for 30 s,
which yields an approximately equal number of atoms of each isotope
during evaporative cooling. The laser parameters for trapping $^{88}$Sr
are given in ref.\ \cite{nsl03}.  For trapping $^{87}$Sr \cite{xlh03},
the laser is approximately 70\,MHz red-detuned from the
$^1S_0(F=9/2)$-$^1P_1(F=11/2)$ transition (slightly more than $2\Gamma$, where $\Gamma=30.5$\,MHz is the natural linewidth of the transition \cite{nms05}).

$^3P_2$ atoms are returned to the ground state using 60~ms of 3 W/cm$^2$
%[intensity calculated from power and beam size from 3 um paper]
of excitation on the $(5s5p)^3P_2$-$(5s4d)^3D_2$ transition at
3\,$\mu$m \cite{mma09}.  The isotope shift ($f_{87}-f_{88}=110$\,MHz) \cite{mma09} is small
compared to the $\sim 500$\,MHz width of the repumping efficiency curve
\cite{mic10} for $^{88}$Sr and the $\sim 3$\,GHz width of the hyperfine
structure  in $^{87}$Sr \cite{mma09}. We tune the 3\,$\mu$m laser 1.6
GHz blue detuned from the $^{88}$Sr resonance, which optimizes the $^{87}$Sr
repumping while reducing the $^{88}$Sr number by 80\%. This is a reasonable
compromise given that the number of $^{87}$Sr atoms is the limiting factor
in the experiment. The 461\,nm MOT is left on at the optimal $^{87}$Sr
detuning to maximize the number of captured $^{87}$Sr atoms, but this is
only $\sim 5$ natural linewidths red-detuned of the $^{88}$Sr
$^1S_0$-$^1P_1$ transition \cite{xld03,klj06}, so it also serves to aid
recapture of this isotope. We typically recapture approximately
$1.1\times 10^7$ $^{88}$Sr  and $3\times 10^7$ $^{87}$Sr at temperatures of a few millikelvin.

The 461\,nm light is then extinguished and 689\,nm light is applied to drive
the $(5s^2)^1S_0$-$(5s5p)^3P_1$ transitions and create intercombination-line MOTs
for each isotope. The resonance frequencies in each isotope are well-resolved
compared to the 7.4\,kHz transition linewidth, so the simultaneous MOTs are
compatible with each other.
The parameters of the $^1S_0$-$^3P_1$ lasers for $^{88}$Sr and $^{87}$Sr \cite{mic10}
are similar to the conditions in ref.\ \cite{kii99} and ref.\ \cite{mki03}
respectively. As many as 70\% of the atoms are initially
captured in the intercombination-line MOT. (For experiments with one isotope, the loading phase and
intercombination-line lasers for the other isotope are omitted.)

After 400\, ms of  $^1S_0$-$^3P_1$ laser cooling, an optical dipole trap (ODT) consisting of
two crossed beams is overlapped for 100 \,ms with the intercombination-line MOT
with modest power (3.9 W) per beam. The ODT is formed by a single beam
derived from a 20 W multimode, 1.06\,$\mu$m fiber laser that is recycled
through the chamber to produce a trap with equipotentials that are nearly
oblate spheroids, with the tight axis close to vertical. Each beam has a
waist of approximately 90 $\mu$m in the trapping region.

Immediately after extinction of the 689 nm light, the ODT  power is
ramped in 30 ms to 7.5 W to obtain a trap depth of 25 $\mu$K. Typically
the atom number, temperature, and peak density at this point for both $^{88}$Sr
and $^{87}$Sr are  $3\times 10^6$, 7\,$\mu$K, and  $2.5 \times 10^{13}$\,cm$^{-3}$. The
peak phase space density (PSD) for $^{88}$Sr is $10^{-2}$.

For diagnostics, we record $^1S_0$-$^1P_1$ resonant absorption
images of samples after a time of flight varying from 10 to 40 \,ms.
Because of broadening of the resonance due to hyperfine structure,
%Hyperfine structure broadens the transition for $^{87}$Sr, so
$^{87}$Sr atoms present would contribute significantly to the absorption when imaging at the $^{88}$Sr resonance frequency. To remove $^{87}$Sr atoms and obtain clean $^{88}$Sr images, light resonant with the $^1S_0(F=9/2)$-$^3P_1(F=11/2)$ transition in $^{87}$Sr is applied during the first 2\,ms of the time of flight. $^{87}$Sr atoms are imaged with linearly polarized light resonant with the $^1S_0(F=9/2)$-$^1P_1(F=11/2)$ transition, and the contamination due to $^{88}$Sr is small and easily accounted for \cite{mic10}.

\begin{figure}
\includegraphics[clip=true,angle=-90,keepaspectratio=true,width=3.3in, trim=0 0 0 0 ]{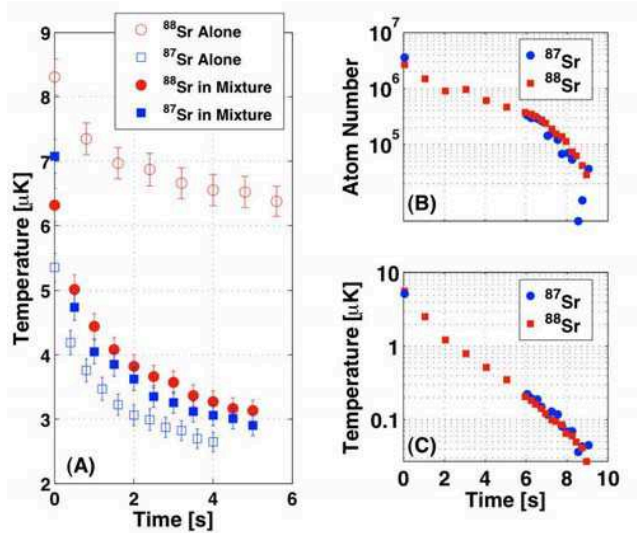}\\
\caption[Optional caption for list of figures]{(color online) (A) Temperature evolution in an ODT with trap depth of $U/k_B=23\,\mu$K for samples of $^{88}$Sr and $^{87}$Sr alone and for each in a mixture. The number of each isotope present initially is approximately $10^6$.  (B) Number and (C) temperature for a mixture along a typical forced evaporation trajectory. }
\label{figure:evaporation}
\end{figure}

To investigate the collisional properties of the different isotopes and the
mixture, the evolution of number and temperature were recorded in a fixed
potential (Fig. \ref{figure:evaporation}A). For $^{88}$Sr alone,
evaporation is inefficient and a typical ratio of the trap depth to the
sample temperature is $\eta\approx 4$, as observed previously \cite{fdp06}.
$^{87}$Sr, however, approaches $\eta\approx 9$.
Modeling \cite{ycm09} of the free-evaporation trajectory for $^{87}$Sr alone suggests a
moderate degree of polarization \cite{mic10} that will be investigated in future studies. The temperatures of
$^{87}$Sr and $^{88}$Sr atoms in a mixture  with peak densities of
$8 \times 10^{12}$\,cm$^{-3}$ track each other closely and approach $\eta\approx 8$, indicating that $^{87}$Sr
provides efficient sympathetic cooling of $^{88}$Sr.

Figure\ \ref{figure:evaporation} shows the number (B) and temperature (C) for  a typical forced evaporation trajectory with a mixture. We decrease
the laser power  according to $P=P_0/(1+t/\tau)^{\beta}+P_{offset}$, with time
denoted by $t$,  $\beta=1.4$, and $\tau=1.5$\,s. This trajectory without $P_{offset}$ was designed \cite{ogg01} to yield efficient evaporation when gravity can be neglected. Gravity is a significant effect in this trap for Sr, and to avoid decreasing the potential depth too quickly at the end of the evaporation, we set $P_{offset}=0.7$\,W, which corresponds to the power at which gravity causes the trap depth to be close to zero.
 The lifetime of atoms in the ODT is 30 s. This allows efficient
evaporation and an increase of PSD by a factor of 100 for a loss of one order of magnitude in the number of atoms.
The $^{87}$Sr and $^{88}$Sr remain in equilibrium with each other during the evaporation. $^{87}$Sr atoms are lost at a slightly faster rate, as expected because
essentially every collision involves an $^{87}$Sr atom.

\begin{figure}
\includegraphics[clip=true,keepaspectratio=true,width=3.3in, trim=0 0 0 0 ]{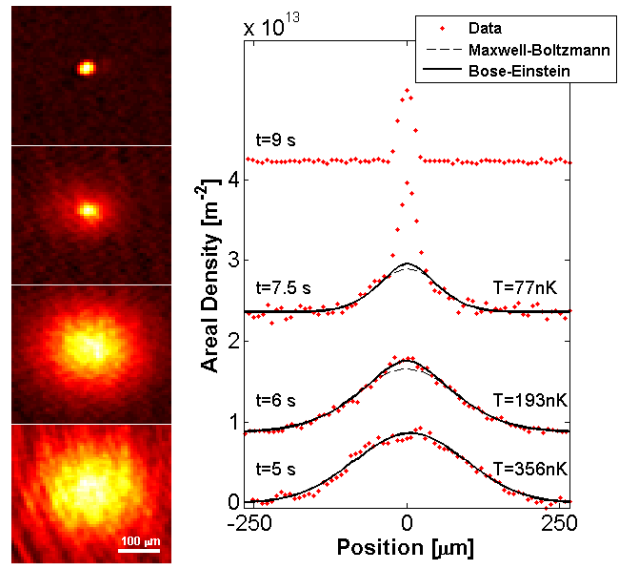}\\
\caption{(color online) Appearance of Bose-Einstein condensation in absorption
images (left) and areal density profiles (right). Data correspond to 16\,ms (bottom) or 22\,ms
(top three) of
free expansion after indicated evaporation times (t).  Images on the left
 have the same time stamp  as on the right.
The areal density profiles are from a vertical cut through the center
of the atom cloud, and temperatures are extracted from 2D Bose-Einstein distribution
fits to the thermal pedestal.
At 7.5\,s, a bimodal distribution is evident and indicative of
Bose-Einstein condensation. For bimodal data,
the central region is excluded from the fits
and the fugacity is constrained to 1.
A pure condensate is shown at 9\,s of evaporation.
} \label{Bimodal Distribution}
\end{figure}

Figure\ \ref{Bimodal Distribution} shows false color 2-dimensional renderings of (left) and
1-dimensional slices through (right) the time-of-flight absorption images recorded after 16\,ms or 22\,ms of
expansion for various points along the evaporation trajectory. At 5\,s of
evaporation, the distribution is fit well by a Boltzmann distribution, but at
6\,s, a Boltzmann distribution fit to the high velocity wings clearly underestimates the number of atoms at low velocity. A fit using the Bose-Einstein distribution  \cite{kds99},
 however, matches the distribution well. The fugacity obtained from this fit is $1.0$,
  indicating  this is close to the critical temperature for condensation.
 With further evaporation,  the presence of a Bose-Einstein condensate is indicated by the emergence
of a narrow peak at low velocity, and eventually, a pure condensate is observed.

At the transition temperature, $2\times10^5$ $^{87}$Sr  atoms remain at a temperature of 0.2 $\mu$K.
This corresponds to ${T}/{T_F}=0.9$ for an unpolarized sample, which is non-degenerate and above the point at which Pauli
blocking significantly impedes evaporation efficiency \cite{dji99}.

$^{88}$Sr has a negative scattering length, so one expects a collapse of the condensate when the system
 approaches a critical number of condensed atoms given by
\cite{rhb95}
\begin{equation}\label{Equation: Maximum BEC number}
    N_{cr}=0.575\frac{a_{ho}}{\left| a_{88} \right|}.
\end{equation}
Here $a_{ho}=[\hbar/(m\overline{\omega})]^{1/2}$ is the harmonic oscillator length, where $m$ is the atom mass, $\hbar$
is the reduced Planck constant, and $\overline{\omega}=(\omega_x\omega_y\omega_z)^{1/3}$ is the geometric average of the oscillator frequencies.
 One should also see large fluctuations in the number of condensed atoms during the evaporation due to repeated collapses
 and refilling of the condensate. To investigate this, we recorded the condensate number for various points in the
 evaporation trajectory over many experimental runs. For absorption images with a condensate and thermal pedastal, we fit the wings of the
 thermal cloud, which are beyond the condensate radii of about 23\,$\mu$m, to a Bose distribution with fugacity set to 1.
 The residuals of the fit represent the condensate atoms which are fit with the standard Thomas-Fermi functional
 form \cite{dgp99,mmy09} to determine their number.

\begin{figure}

\includegraphics[keepaspectratio=true,width=3.3in,height=2.7in]{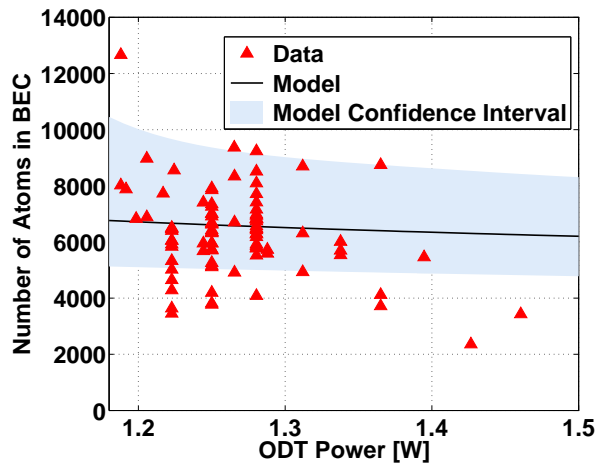}\\
\caption{(color online) Comparison of observed condensate number and maximum condensate number  predicted by a model of the trapping potential
 along the evaporation trajectory.} \label{Limited BEC Number}

\end{figure}

Figure\ \ref{Limited BEC Number} shows the observed condensate number along
the evaporation trajectory from 7 to 10 seconds, as well as maximum
values $N_{cr}$ predicted by Eq.\ \ref{Equation: Maximum BEC number} where $\overline{\omega}$ is determined from
knowledge of the ODT potential,
 with confidence intervals reflecting uncertainties of the waists of ODT beams, of about 10\%, and the uncertainties
 of the scattering length $a_{88} = -2.0(3)$\,$a_0$ \cite{skt10}. As expected for attractive interactions, large fluctuations in number are observed.
While the best guess curve for $N_{cr}$ falls below some of the data, the upper bound is reasonably well accommodated by the confidence interval.
Uncertainties in knowledge of the trap become larger at low ODT power because of the increasingly important
role of gravity, which weakens the trap.

We have described the Bose-Einstein condensation of $^{88}$Sr through sympathetic cooling with $^{87}$Sr.
Observation of large fluctuations in number below a maximum number of condensed atoms is consistent with the small,
negative value of $a_{88}$. Because of the very weak interactions,
it should be possible to change the sign of the scattering length with
an optical Feshbach resonance \cite{ctj05,ekk08} while keeping induced inelastic losses low.
This suggests many possible future experiments, such as creation of matter-wave solitons in two
dimensions \cite{sue03} and  quantum fluids with random nonlinear interactions
\cite{fwg89}.

\textmd{\textbf{Acknowledgements}}
 This research was supported by the Welch Foundation
(C-1579), National Science Foundation (PHY-0855642), and the Keck Foundation.

%\bibliography{bibliography}

\end{document}